%% file: main.tex
\documentclass{ifacconf}
\usepackage{url}
\usepackage{graphicx}      
\usepackage{natbib}        

\usepackage{amsfonts}
\usepackage{bm}
\usepackage{caption}
\usepackage{subcaption}

\usepackage[acronym]{glossaries}
\newacronym{LiL}{LiL}{Live-In Lab}
\newacronym{HVAC}{HVAC}{Heating, Ventilation, and Air Conditioning}
\newacronym[\glsshortpluralkey={CPHS}]{CPHS}{CPHS}{Cyber-Physical-Human System}
\newacronym{BES}{BES}{Building Energy Simulation}
\newacronym{DHI}{DHI}{Diffuse Horizontal Irradiance}

\newcommand{\degrees}{^\circ\mathrm{C}}
\newcommand{\Real}{\mathbb{R}}
\newcommand{\Lset}{\mathcal{L}}
\begin{document}
\begin{frontmatter}

\title{Sustainable Heating with Karma: A Simulation Study of the KTH Live-In Lab\thanksref{footnoteinfo}} 

\thanks[footnoteinfo]{Corresponding author: Ezzat Elokda, elokda@kth.se. This work was supported by the Wallenberg AI, Autonomous Systems and Software Program (WASP) funded by the Knut and Alice Wallenberg Foundation, by the Swedish Energy Authority and IQ Samhällsbyggnad project DOCENT (project number P2023-01513, agreement 2023-205321), and by the Digital Futures project HiSSx.}

\author[First]{Mahsa Farjadnia} 
\author[Second]{Ezzat Elokda} 
\author[Second]{Angela Fontan}
\author[First]{Marco Molinari}
\address[First]{Department of Energy Technology,}
\address[Second]{Department of Decision and Control Systems,\\ KTH Royal Institute of Technology, 100 44 Stockholm, Sweden,\\
E-mail:  \{mahsafa,elokda,angfon,marcomo\}@kth.se.}

\begin{abstract}                
Space heating in buildings accounts for 10\% of the global CO$_2$ footprint.
The widespread adoption of energy-efficient heating technology, e.g., heat pumps, could help reduce this figure, but technology alone may not suffice to reach carbon neutrality.
Additionally, human occupants have an important role to play by adopting sustainable heating behaviors, e.g., avoid excessive window opening in the winter or (pre-)heat their units while clean energy is abundant. Thus far demand response policies aimed at promoting these behaviors have been monetary, which discriminates against low-income households and exposes human occupants who do not actively engage with real-time control signals to financial risks.
This paper instead investigates the suitability of a non-monetary karma economy for promoting sustainable heating behaviors.
Karma leverages the repeated and dynamic nature of heating energy allocations to attain climate targets both fairly and efficiently over time without resorting to financial means.
As a first step towards experimentally validating the karma concept with real human occupants in the KTH Live-In Lab, we perform a simulation study on a digital model of the Live-In Lab.
The study provides initial estimates of expected effects to guide the design of human-in-the-loop experiments, as well as assists with designing and tuning the karma economy in this context.
As a specific example, we investigate how incorporating consumption memory in the form of karma affects window opening behaviors in comparison to conventional memory-less heating operation.
\end{abstract}


\end{frontmatter}

\input{sections/introduction}
\input{sections/setup}
\input{sections/karma}
\input{sections/results}
\input{sections/conclusion}



\bibliography{main}             
                                                   







\appendix
\input{sections/bidding-policy}

\end{document}

%% file: sections/introduction.tex
\section{Introduction}

Space heating accounts for roughly 10\% of global greenhouse gas emissions \citep{IEA_Building}.
Occupant behavior is a key determinant of building energy performance, especially in space heating operation \citep{fabi2012occupants,xu2023critical,farjadnia2026assessing}. This has motivated viewing modern buildings as \glspl{CPHS}: an emerging interdisciplinary area concerned with the dynamic interactions between cyber-physical systems, physical systems integrated with computational devices, and humans \citep[Chapter~4]{annaswamy2024control}.

In particular, \emph{demand response} has emerged as a promising policy instrument to incentivize occupants to adopt sustainable energy consumption behaviors, i.e., shift or limit their consumption patterns to align with the availability of clean energy and/or provide needed flexibility in renewable-dominated power grids~\citep{dedecca2025increasing,kondziella2016flexibility}.
The essence of demand response is to expose occupants to dynamic, real-time price signals that are high during energy scarcity, thereby incentivizing off-peak consumption.
However, the majority of proposed demand response schemes have been monetary~\citep{grvzanic2022prosumers}, which a) can be highly \emph{discriminatory}, shifting consumption based on willingness to pay while overlooking ability to pay~\citep{palm2026americans}; and b) expose occupants who don't carefully monitor and react to real-time prices to financial risks.
This paper instead investigates a \emph{non-monetary} approach for promoting sustainable consumption behaviors based on the recently developed concept of \emph{karma economies}~\citep{elokda2024self,elokda2025carma,elokda2025vision,pedroso2024fair}.
Karma is a \emph{non-tradable} token used to bid for resources repeatedly, flowing from resource consumers to yielders in a closed and indefinitely sustainable cycle.
This design encourages truthful bidding according to private needs, as users effectively ``play against their future selves.''
Thus, rather than charging occupants money for their behaviors, karma charges in ``future comfort'' as it essentially budgets consumption over time.

In comparison to previous works, which focused on formalizing karma economies as games~\citep{elokda2024self}, showing their theoretical fairness and efficiency properties~\citep{elokda2025vision}, and behavioral validation in online experiments~\citep{elokda2024dynamic}, this paper forms a stepping stone towards the first real-world implementation of the karma concept.
Specifically, we consider the problem of allocating a limited budget of energy to heat units of the KTH Live-In Lab\footnote{\url{https://www.liveinlab.kth.se/en}.}, an experimental facility housing real occupants located on the KTH campus in Stockholm, Sweden.
This energy budget may correspond to locally sourced solar energy, or decreased reliance on grid energy when real-time prices are high; and the goal is to leverage the building's thermal inertia as well as occupants' time-varying needs for space heating to fairly and efficiently allocate this budget.
A karma economy in this context is expected to incentivize sustainable occupant behaviors such as avoiding excessive window opening during the cold winter months or decreasing heating setpoints when occupants are not present in their units.

We perform a realistic simulation study on a digital model of \emph{Testbed KTH}, a four-unit apartment within the KTH Live-In Lab commonly utilized for pilot experiments.
The purpose of the simulation study is to get an initial estimate of expected effects in order to guide the design of real human-in-the-loop experiments; and develop an understanding of how to tune parameters of the karma economy, e.g., how to redistribute karma payments to each unit after each time-step.
The main contributions of this paper are twofold: to develop a co-simulation platform that interfaces between the karma economy and the units' temperature controllers; and to provide initial simulation results in support of the above goals.
To summarize the simulation results, we find that the karma scheme is indeed capable of disincentivizing unsustainable behaviors in the form of excessive window opening; and is effective in reducing the apartment's space heating energy footprint despite the presence of model mismatches.
However, effective parameter tuning requires simulating the karma scheme during nominal symmetric conditions, cf. Section~\ref{subsec:params}, which may be challenging in practice and points to the importance of well-calibrated digital models.

The paper is organized as follows: Section~\ref{sec:setup} provides an overview of the building's physical and digital representations. Section~\ref{sec:karma} introduces the karma-based scheme for allocating space heating energy to Testbed KTH units by adjusting the setpoint temperatures for each unit. Section~\ref{sec:results} presents the simulation results. Finally, Section~\ref{sec:conclusion} offers conclusive remarks and discusses future directions.

%% file: sections/setup.tex
\section{Experimental Setup}\label{sec:setup} 

In this section, we first present Testbed KTH, followed by IDA ICE, the simulation tool through which a digital model of the building was constructed. Finally, we outline the coupled IDA ICE–MATLAB co-simulation framework, which evaluates the impact of karma-based allocations.

\begin{figure}[h!]\centering
\subfloat[]{\includegraphics[height=3.8cm,trim={0 0 0 0.5cm},clip]{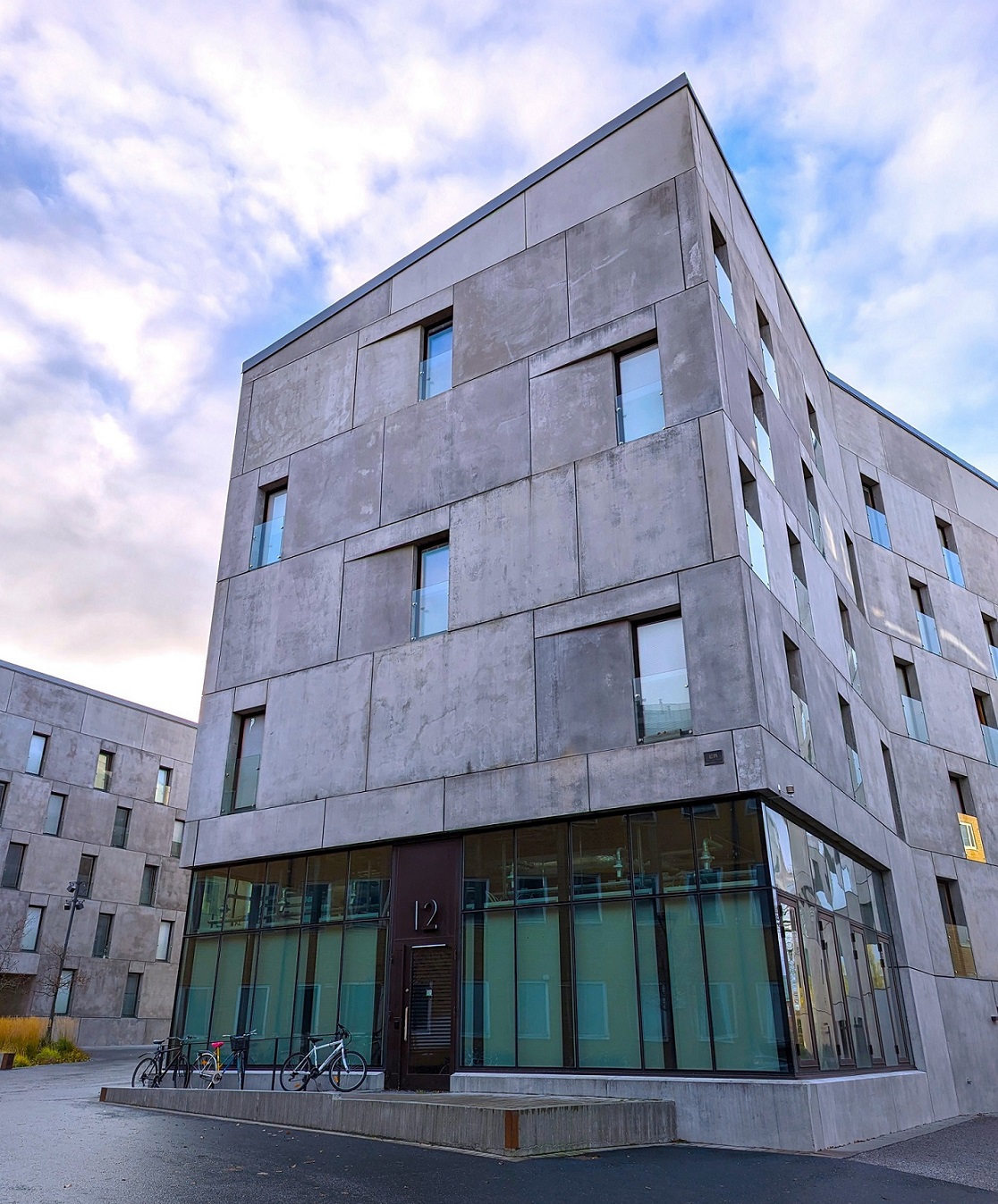}\label{fig:Outside}}
\quad \quad
\subfloat[]{\includegraphics[height=4cm]{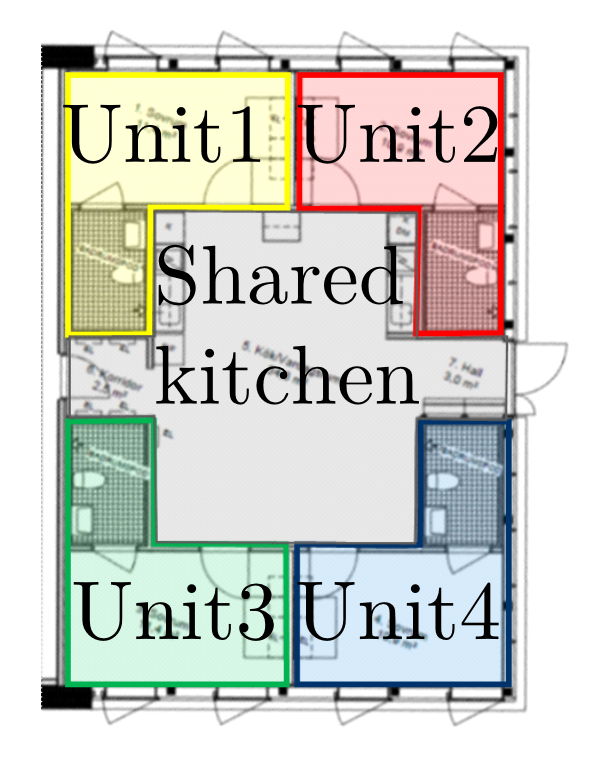}\label{fig:Testbed2}}
\caption{(a): Exterior view of Testbed KTH. (b): Testbed KTH layout (August 2020 to June 2021).}
\label{fig:TestbedEM}
\end{figure}

    The KTH Live-In Lab comprises a range of building testbeds, from student housing to lecture halls. This study focuses on Testbed KTH (Fig.~\ref{fig:Outside}), a residential apartment that houses students full-time and features a reconfigurable layout of roughly 300$\mathrm{m}^2$, divided into four units with a shared kitchen (Fig.~\ref{fig:Testbed2}).
We adopt Layout 2.0, which was in use between August 2020 and June 2021.

\begin{figure}[h!]\centering
\includegraphics[height=4.1cm]{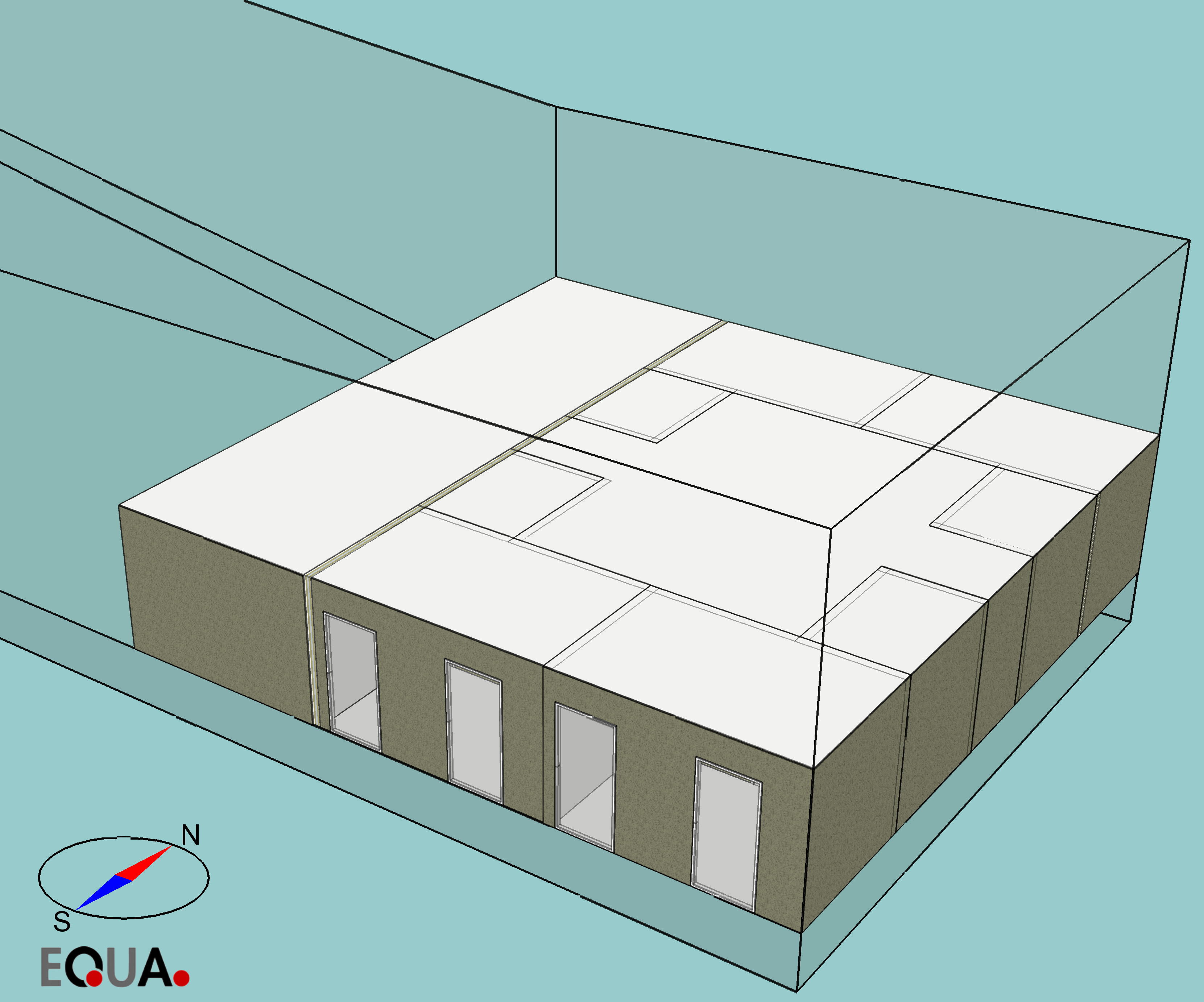}
\caption{Digital model of Testbed KTH in IDA ICE.}
\label{fig:IDAICE}
\end{figure}

A calibrated digital model of Testbed KTH was developed in IDA ICE~\citep{kalamees2004ida}, a \gls{BES} tool for simulating building thermal behavior, air quality, and energy use, cf. Fig.~\ref{fig:IDAICE} for an illustration and \cite{farjadnia2026assessing} for further details. 
However, \gls{BES} programs are typically not directly suitable for the design and testing of advanced controllers~\citep{drgovna2020all}. To address this issue, a co-simulation environment is designed to establish communication bridges between the \gls{BES} and control-oriented tools and programming languages, such as MATLAB. During runtime, IDA ICE exports selected environmental variables, including indoor and outdoor temperatures, window states, \gls{DHI}, and space heating power, at each IDA ICE/MATLAB synchronization step.
A karma-based scheme, described in Section~\ref{sec:karma}, allocates space heating energy for each unit in response to simulated environmental conditions and occupants' desired setpoints in MATLAB. This is performed by computing and sending adjusted setpoints back to IDA ICE. To ensure stable co-simulation, a synchronization step of 7.5 minutes is used, while the karma-based scheme runs and issues setpoint commands only every 30 minutes. It is assumed that occupants are always present during the simulation time.

%% file: sections/karma.tex
\section{Karma-based Space Heating Allocation}\label{sec:karma} 

\begin{figure}[!t]
    \centering
    \includegraphics[width=0.8\linewidth]{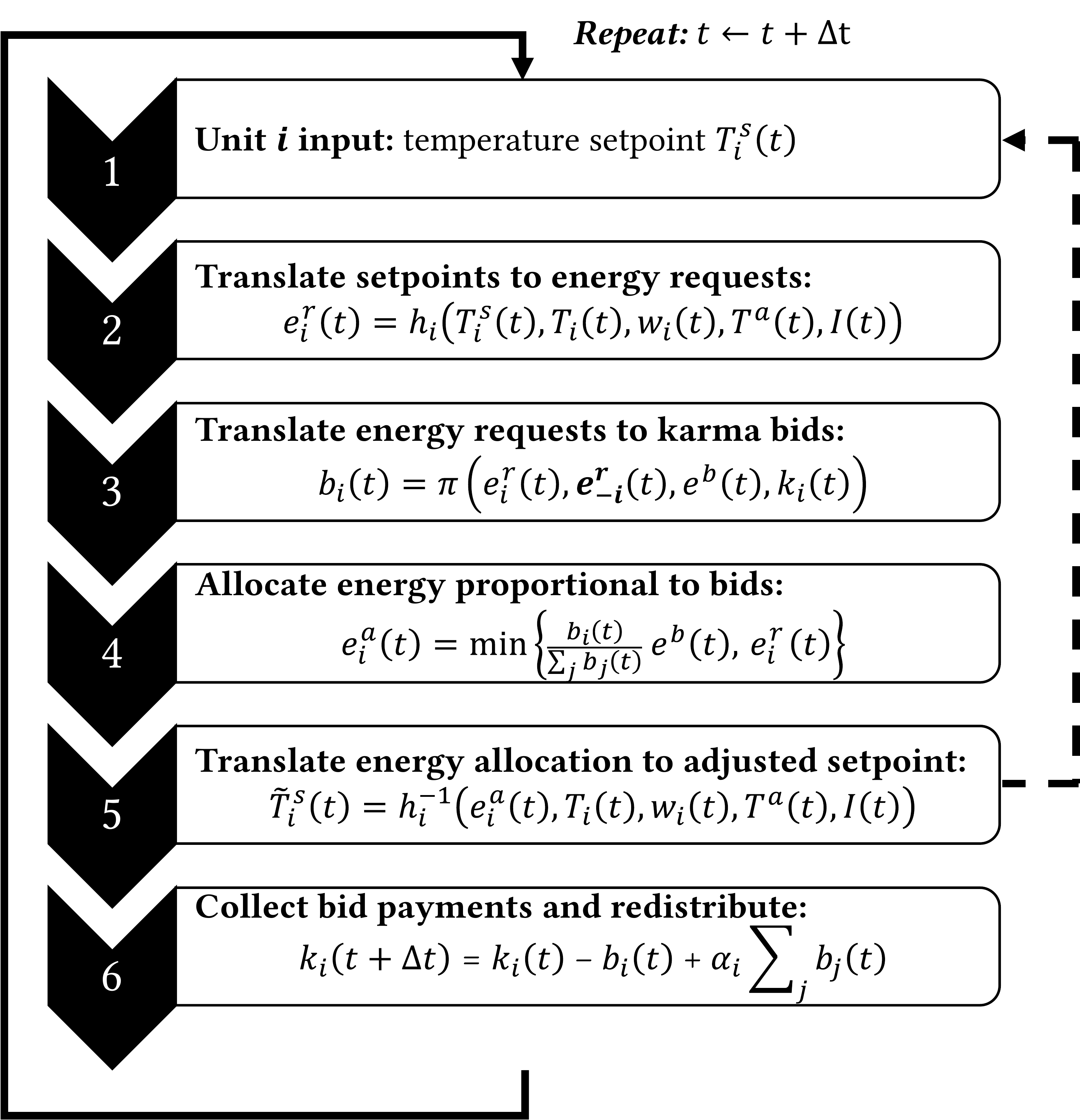}
    \caption{Karma-based heating energy allocation scheme}
    \label{fig:karma-procedure}
\end{figure}

This section describes the karma-based scheme for allocating space heating energy to the Testbed KTH units.
The scheme runs recurrently at discrete time-steps $t \in \{0,\Delta t, 2\Delta t,\dots\}$ and allocates a (potentially time-varying) total energy budget $e^b(t) \in \Real_+$ to heat each unit in the forthcoming $\Delta t=30$min time interval.
Each unit $i \in \{1,\dots,4\}$ is endowed with a time-varying budget of karma $k_i(t) \in \Real_+$ which encodes the unit's consumption history, and initially all units get $k_i(0) = \bar k = 10$.
The six-step procedure illustrated in Fig.~\ref{fig:karma-procedure} is then repeated in each time-step $t$, as elaborated below.

\emph{Step 1: Unit Input Setpoint.}
The system probes the current temperature setpoint $T_i^s(t)$ of each unit $i$.

\emph{Step 2: Setpoints to Energy Requests.}
A first-order heat transfer model predicts space heating energy requests $e^r_i(t)$ required for each unit $i$ to reach and/or maintain its setpoint $T_i^s(t)$ over the next time interval $\Delta t$, based on the unit's current temperature $T_i(t)$, window opening state $w_i(t) \in [0,1]$ ($w_i(t)=0$: fully closed; $w_i(t)=1$: fully open), the ambient temperature $T^a(t)$, and \gls{DHI} $I(t)$:
\begin{subequations}
\label{eq:setpoint-to-request}
\begin{align}
    e^r_i(t) &= h_i(T_i^s(t), T_i(t), w_i(t), T^a(t), I(t)) \\
    &= c_i^\textup{tr} \left(T_i^s - T_i\right) + c_i^\textup{l}(w_i) \left(T_i^s - T^a\right) - c_i^\textup{ir} \, I, 
\end{align}
\end{subequations}
where $c^\textup{tr}_i \in \Real_+$ captures transient effects, $c^\textup{irr}_i \in \Real_+$ is the heat gain coefficient due to \gls{DHI}, and
\begin{align}
    c^\textup{l}_i(w_i) = c^\textup{l,wall}_i + (1 - w_i) \, c^\textup{l,win-cl}_i + w_i \, c^\textup{l,win-op}_i
\end{align}
is the heat loss coefficient due to the ambient temperature, which depends on the extent of window opening: $c^\textup{l,win-cl}_i$ (respectively, $c^\textup{l,win-op}_i > c^\textup{l,win-cl}_i$) is the fully closed (respectively, fully open) window heat loss coefficient, and $c^\textup{l,wall}_i,c^\textup{l,win-cl}_i,c^\textup{l,win-op}_i \in \Real_+$.
Note that this model makes several simplifying assumptions, including approximating transient effects with a first-order term and assuming constant ambient temperature, window opening, and solar irradiation over $\Delta t$. Its purpose is not model realism but to provide rough estimates for placing karma bids.

\emph{Step 3: Energy Requests to Karma Bids.}
The system then places karma bids on the units' behalf according to the following heuristic bidding policy
\begin{subequations}
\label{eq:bidding-policy}
\begin{align}
    b_i(t) &= \pi(e^r_i(t), \bm{e^r_{-i}}(t), e^b(t), k_i(t)) \\
    &= \min\left\{\beta \left[1 - \frac{e^b}{\sum_j e^r_j}\right]_+ e^r_i, \, k_i\right\},
\end{align}
\end{subequations}
where $\bm{e^r_{-i}}(t)$ denotes the concatenated vector of energy requests from units other than $i$, and we use the notation $\left[\cdot\right]_+:=\max\{\cdot,0\}$.
The structure of this heuristic policy is loosely guided by the analysis
in~\cite{elokda2025vision}, cf. Appendix~\ref{sec:bidding-policy}.
Namely, unit $i$'s bid is constrained by its current karma budget $k_i(t)$; otherwise,
it is linear in the energy request $e^r_i(t)$, and increases as the competition gets more severe, i.e., as the supply-to-demand ratio $e^b / \sum_j e^r_j$ decreases.
In case there is no competition ($e^b / \sum_j e^r_j \geq 1$), bids default to zero.
The scaling parameter $\beta$ controls how quickly karma budgets deplete.
Tuning this parameter forms a trade-off: to avoid premature budget depletion, $\beta$ should be sufficiently small such that $b_i(t) \ll \bar k$; however, smaller values allow unsustainable behaviors to persist for longer before they are ``punished'' by the lack of karma.

\emph{Step 4: Karma Bids to Energy Allocations.}
In case there is no supply shortage, i.e., $\sum_i e^r_i(t) \leq e^b(t)$, then each unit simply gets allocated its energy request, i.e., $e^a_i(t) = e^r_i(t)$, and recall the karma bids default to zero leading to no exchange of karma.
Otherwise, heating energy is allocated proportionally to the bids, up to the energy requested, i.e.,
\begin{align}
    e^a_i(t) = \min\left\{\frac{b_i(t)}{\sum_j b_j(t)} \, e^b(t), \, e^r_i(t)\right\}.
\end{align}
This proportional allocation rule is standard in the literature on fair division~\citep{kelly1998rate,moulin2004fair,nisan2007algorithmic}.
Bids accordingly represent \emph{access rights} or \emph{entitlements}, and the available energy budget is allocated proportionally to these entitlements.

During regular operation, when $\beta \left[1 - \frac{e^b(t)}{\sum_j e^r_j(t)}\right]_+ e^r_i(t) \leq k_i(t)$ for all units (i.e., none of the karma budget constraints bind), it is straightforward to show that $e^a_i(t) = \frac{e^r_i(t)}{\sum_j e^r_j(t)} \, e^b < e^r_i(t)$, i.e., energy is allocated proportionally to the requests.
However, this ceases to be the case when one or more units' karma budget constraint binds, and let $\Lset \subseteq \{1,\dots,4\}$ denote the set of those units.
For $i \in \Lset$, it will hold that $e^a_i(t) < \frac{e^r_i(t)}{\sum_j e^r_j(t)} \, e^b$, and vice versa for $i \notin \Lset$.
In case, as a result, $\frac{b_i(t)}{\sum_j b_j(t)} \, e^b(t) > e^r_i(t)$ for some $i \notin \Lset$, the allocation is capped at $e^r_i(t)$, and the resulting excess available energy $\frac{b_i(t)}{\sum_j b_j(t)} \, e^b(t) - e^r_i(t)$ is further allocated to the other units proportionally to their bids.
This, in turn, could lead to some other unit $j \notin \Lset$ to cap at its request, and this process is repeated until all of the available energy budget is fully allocated.

\emph{Step 5: Energy Allocations to Adjusted Setpoints.}
Since the IDA ICE temperature controllers take as input temperature setpoints rather than energy or power limits, the energy allocations $e^a_i(t)$ are translated back to adjusted setpoints $\tilde T^s_i(t)$ by inverting Equation~\eqref{eq:setpoint-to-request}, i.e.,
\begin{subequations}
\label{eq:allocation-to-setpoint}
\begin{align}
    \tilde T^s_i(t) &= h^{-1}_i(e^a_i(t), T_i(t), w_i(t), T^a(t), I(t)) \\
    &= \frac{e^a_i + c_i^\textup{tr} \, T_i + c_i^\textup{l}(w_i) \, T^a + c_i^\textup{ir} \, I}{c_i^\textup{tr} + c_i^\textup{l}(w_i)}.
\end{align}
\end{subequations}

\emph{Step 6: Settle Karma Payments and Redistribution.}
The final step is to settle the karma payments and redistribute the total payment.
Each unit pays its karma bid, and the total payment gets redistributed according to unit-dependent shares $\alpha_i \in \Real_+$, $\sum_i \alpha_i = 1$, leading to
\begin{align}
    \label{eq:karma-dynamics}
    k_i(t + \Delta t) &= k_i(t) - b_i(t) + \alpha_i \, \sum_j b_j(t).
\end{align}
The redistribution shares $\alpha_i$ account for persistent heterogeneity in the units, e.g., whether some units persistently require more or less heating because they are North or South-facing, have different average occupancy, etc..
In~\cite{elokda2025vision}, it is shown that these shares correspond to exogenous priority weights or access rights.
In practice, however, it is unclear how to determine these shares a-priori.
Our approach is to tune these shares based on simulations of nominal symmetric conditions, in which all units behave symmetrically in the sense that they follow the same temperature setpoints and window opening actions.
Under such symmetric conditions, the shares are set in a manner that makes each unit's average karma payment equal its average redistribution.

%% file: sections/results.tex
\section{Simulation Results} \label{sec:results} 

This section evaluates the impact of the proposed karma-based allocation scheme on space heating energy consumption and occupants' discomfort level with respect to their desired temperature setpoints.
In order to highlight the effects of the karma scheme, we compare it to a memory-less proportional allocation baseline under two levels of energy scarcity.
Before presenting our simulation results in Section~\ref{subsec:sim-results}, Section~\ref{subsec:params} details the parameter settings used in the simulations, and Section~\ref{subsec:configs} describes the experimental configurations considered.

\subsection{Parameter Settings}
\label{subsec:params}
\begin{figure}[t]
    \centering
    \includegraphics[width=0.8\linewidth]{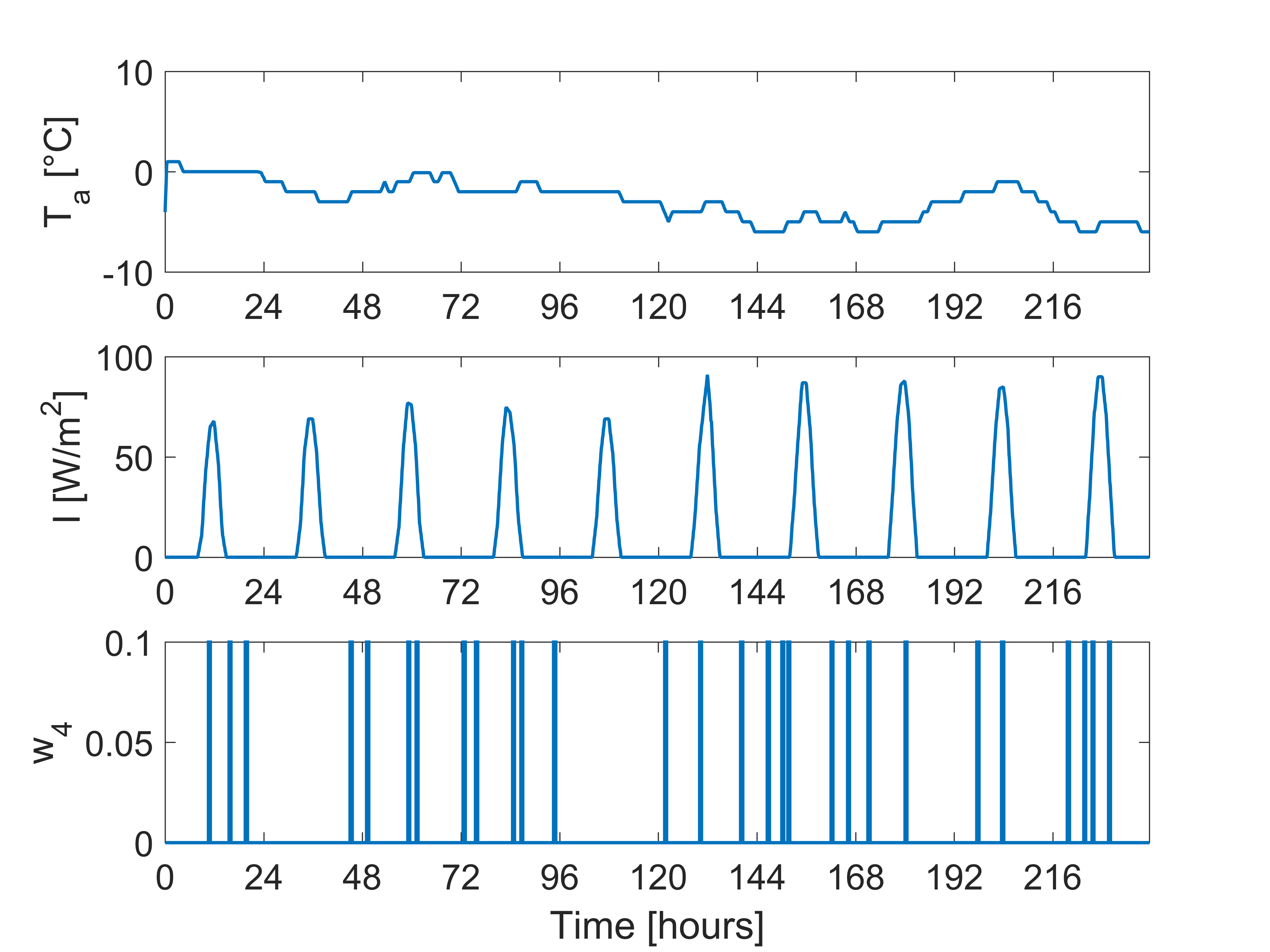}
    \caption{Ambient temperature $T^a(t)$; \gls{DHI} $I(t)$; and window opening state $w_4(t)$ of Unit 4.}
    \label{fig:environemtalCond}
\end{figure}

We keep temperature setpoints fixed and symmetric at $T^s_i(t)=22\degrees$ for all units and time-steps.
We introduce an asymmetry in window opening behaviors, whereby windows remain closed in all units except Unit 4, in which the window is opened approximately twice per day at randomly selected times (bottom panel of Fig.~\ref{fig:environemtalCond}).
This window-opening pattern mimics real historical data observed in \cite{farjadnia2026assessing}, which also suggests that most window openings last less than an hour at roughly 10$\%$ opening angle.
The simulation period spans ten days, and we use averaged real-world data for the ambient temperature $T^a(t)$ and \gls{DHI} $I(t)$ from Stockholm in the period 15-24 January (upper two panels of Fig.~\ref{fig:environemtalCond}).

The parameters of the heat transfer model used in the karma scheme (Eq.~\eqref{eq:setpoint-to-request} and \eqref{eq:allocation-to-setpoint}) were estimated using crude gray-box modeling, combining both known building parameters and loosely fitting them to simulated observations.
On the other hand, the parameters of the karma economy, namely, scaling coefficient $\beta$ in Eq.~\eqref{eq:bidding-policy} and the redistribution shares $\alpha_i$ in Eq.~\eqref{eq:karma-dynamics}, were tuned in a simulation in which Unit 4's window was also kept closed (in addition to setpoints being symmetric), in order to ensure that karma budgets do not deplete prematurely (as controlled by $\beta$); and each unit's karma expenditures roughly equal its gains from redistribution (as controlled by $\alpha_i$) in this symmetric setting.

\subsection{Experimental Configurations}
\label{subsec:configs}
We consider the following configurations.

\noindent\textbf{Configuration 1 [Low Scarcity]:} In this configuration, we consider a fixed energy budget $e^b(t) =375$Wh for all time-steps $t$, which approximately equals the average aggregate space heating energy consumption when all windows remain closed in the absence of budget limits.
We run this configuration with each of the following two space heating energy allocation schemes.
\begin{itemize}
    \item[(1.1)] \emph{Proportional allocation baseline:} The baseline scheme mirrors Steps 1, 2, and 5 of the karma-based scheme, cf. Section~\ref{sec:karma}, but differs in how the energy budget is allocated as a function of the units' energy requests.
    Similarly to the karma scheme, if there is no supply shortage, i.e., $\sum_i e^r_i(t) \leq e^b(t)$, then each unit receives its full request, i.e., $e^a_i(t) = e^r_i(t)$.
    Otherwise, the available budget is allocated proportionally to the requests, $e^a_i(t) = \frac{e^r_i}{\sum_j e^r_j} \, e^b(t)$.
    As discussed in Section~\ref{sec:karma}, Step~4, this allocation coincides with that of karma as long as karma budget constraints do not bind.
    The main distinction is thus that the baseline is \emph{memory-less}, i.e., it does not consider the history of allocations in the present one.

    \item[(1.2)] \emph{Karma-based allocation:} The space heating energy allocation follows the proposed karma-based scheme.
\end{itemize}  

\noindent\textbf{Configuration 2 [High Scarcity]:}
In this configuration, we lower the fixed energy budget to $e^b(t) =300$Wh at all time-steps $t$, which is in fact insufficient to maintain the desired temperature setpoints of $22\degrees$ (even when all windows are closed).
This creates a more constrained scenario in which the aggregate heating request exceeds the budget more frequently.
As before, we run this configuration with (2.1) the proportional allocation baseline and (2.2) the proposed karma-based allocation scheme.

\subsection{Results}
\label{subsec:sim-results}
\begin{figure}[t]
    \centering
    \includegraphics[width=0.8\linewidth]{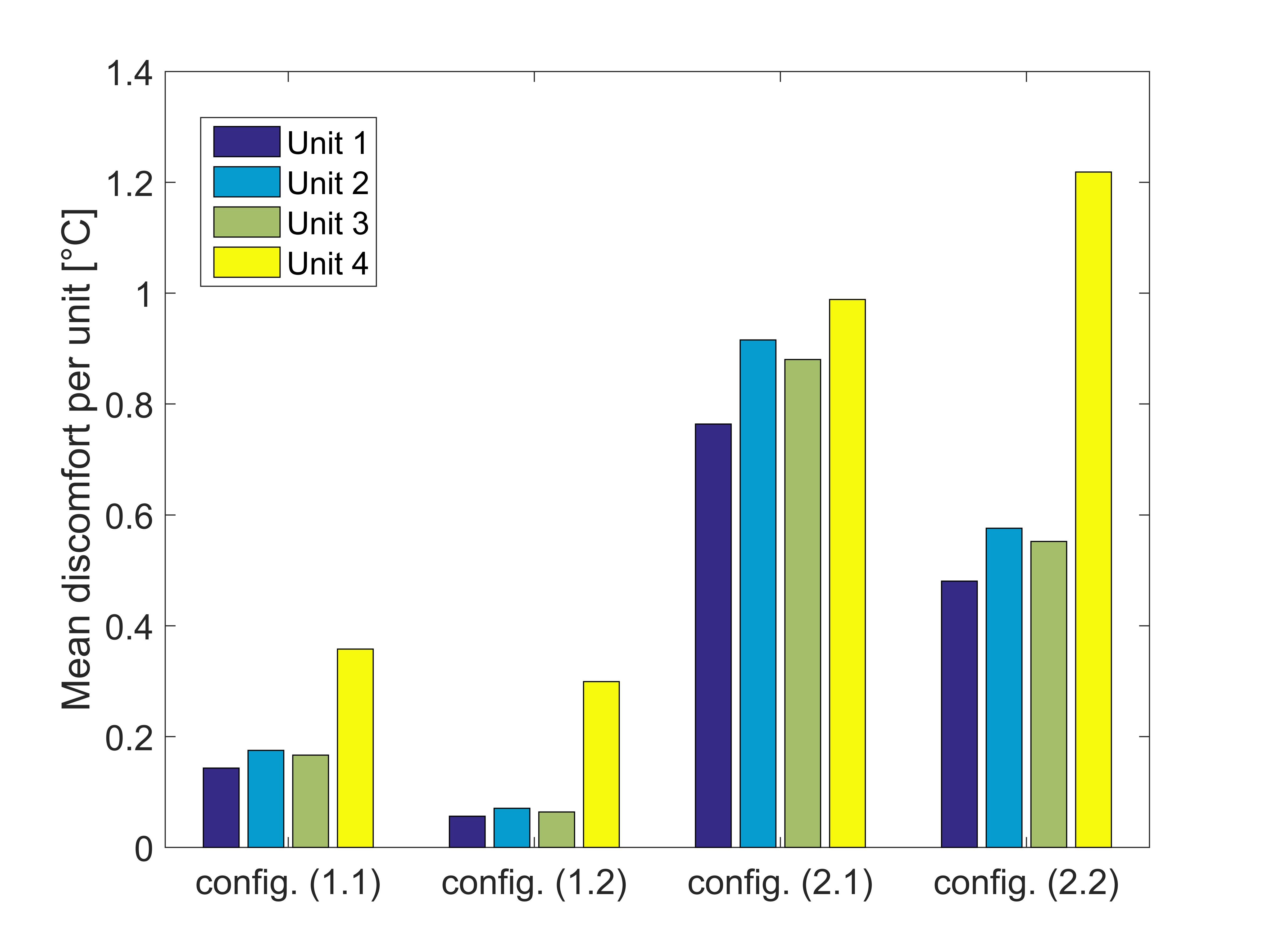}
    \caption{Discomfort level under low- and high-scarcity configurations for proportional allocation baseline ((1.1), (2.1)) and karma-based allocation ((1.2), (2.2)).}
    \label{fig:discomfort}
\end{figure}

Fig.~\ref{fig:discomfort} presents the average discomfort level defined as the mean value of $|T^s_i-T_i|$ for each unit under the considered configurations.
An important observation is that under the proportional allocation baseline, all occupants experience increased discomfort as a consequence of the relatively unsustainable window-opening behavior of Unit 4, where repeated window-opening events increase the aggregate heating demand.
In contrast, the karma-based allocation effectively ``decouples'' the units in the sense that Unit 4 bears the consequences of its own window-opening actions: this unit experiences significantly higher discomfort than the other units, which are relatively unaffected.

\begin{figure*}[t]
    \centering
    \begin{subfigure}[b]{0.49\linewidth}
        \centering
        \includegraphics[width=0.8\linewidth]{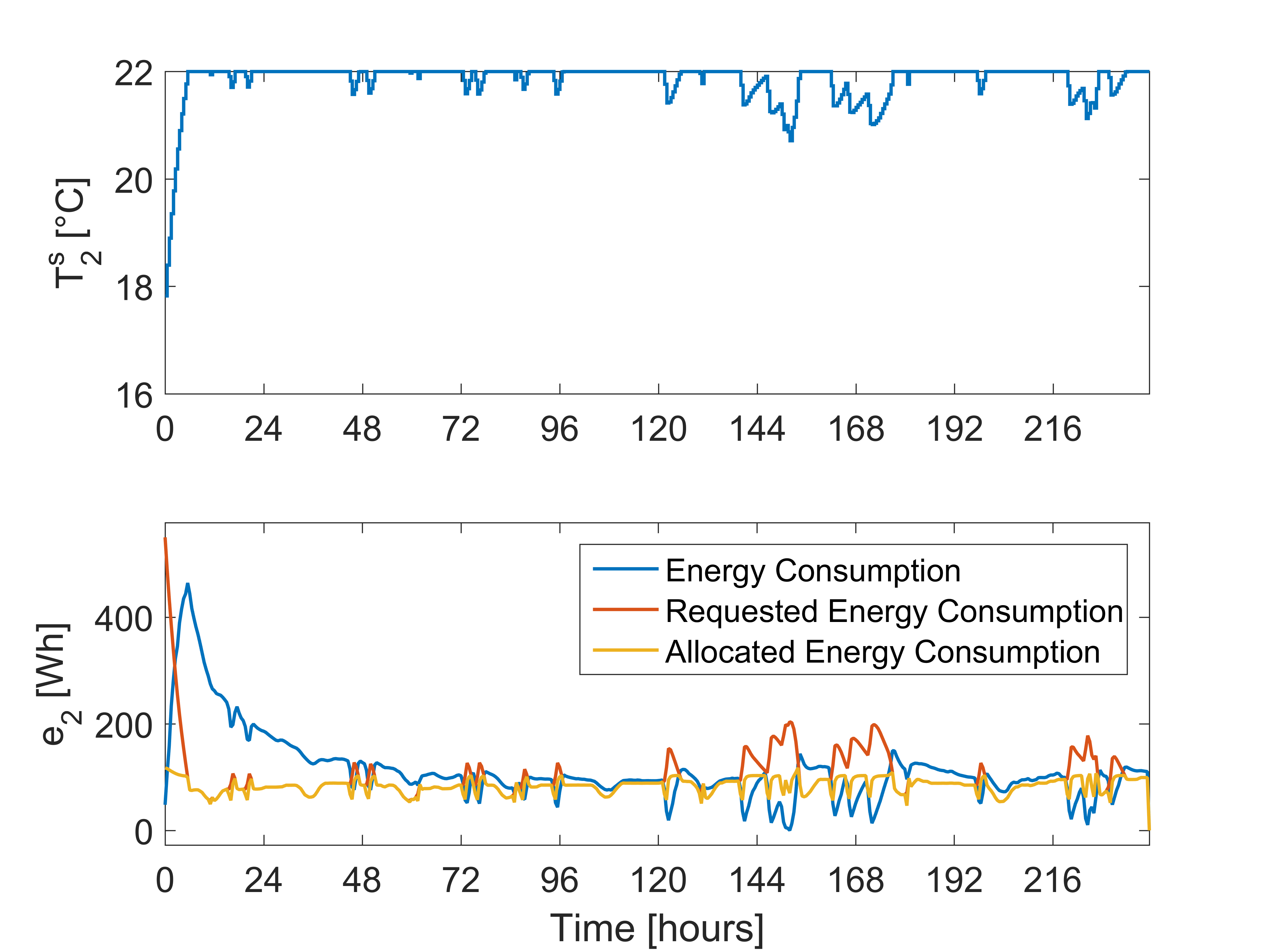}
        \caption{Proportional Allocation, Unit 2}
        \label{fig:TsE_Z2_Karma_750}
    \end{subfigure}
    \hfil
    \begin{subfigure}[b]{0.49\linewidth}
        \centering
        \includegraphics[width=0.8\linewidth]{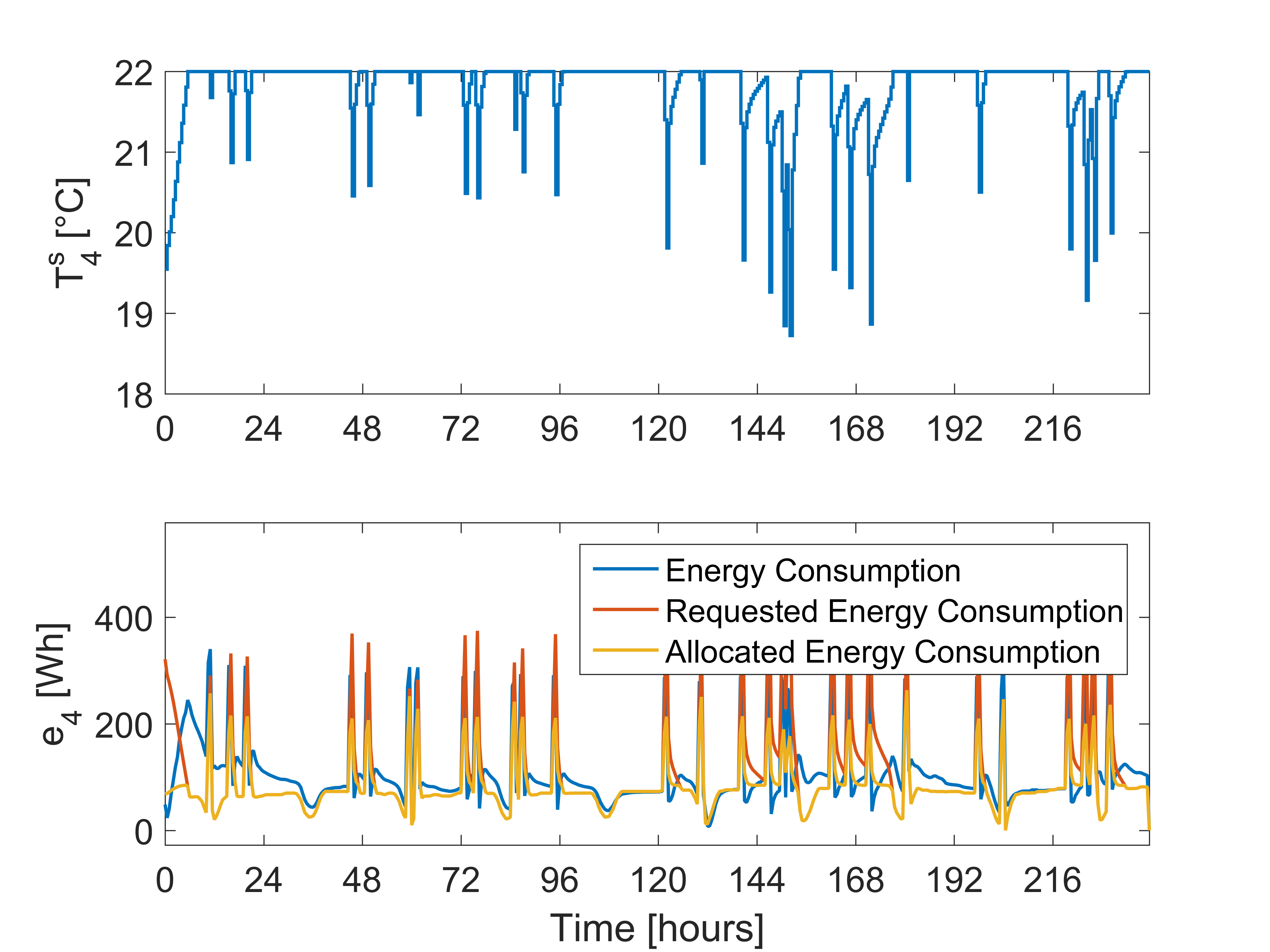}
        \caption{Proportional Allocation, Unit 4}
        \label{fig:TsE_Z4_Prop_750}
    \end{subfigure}

    \begin{subfigure}[b]{0.49\linewidth}
        \centering
        \includegraphics[width=0.8\linewidth]{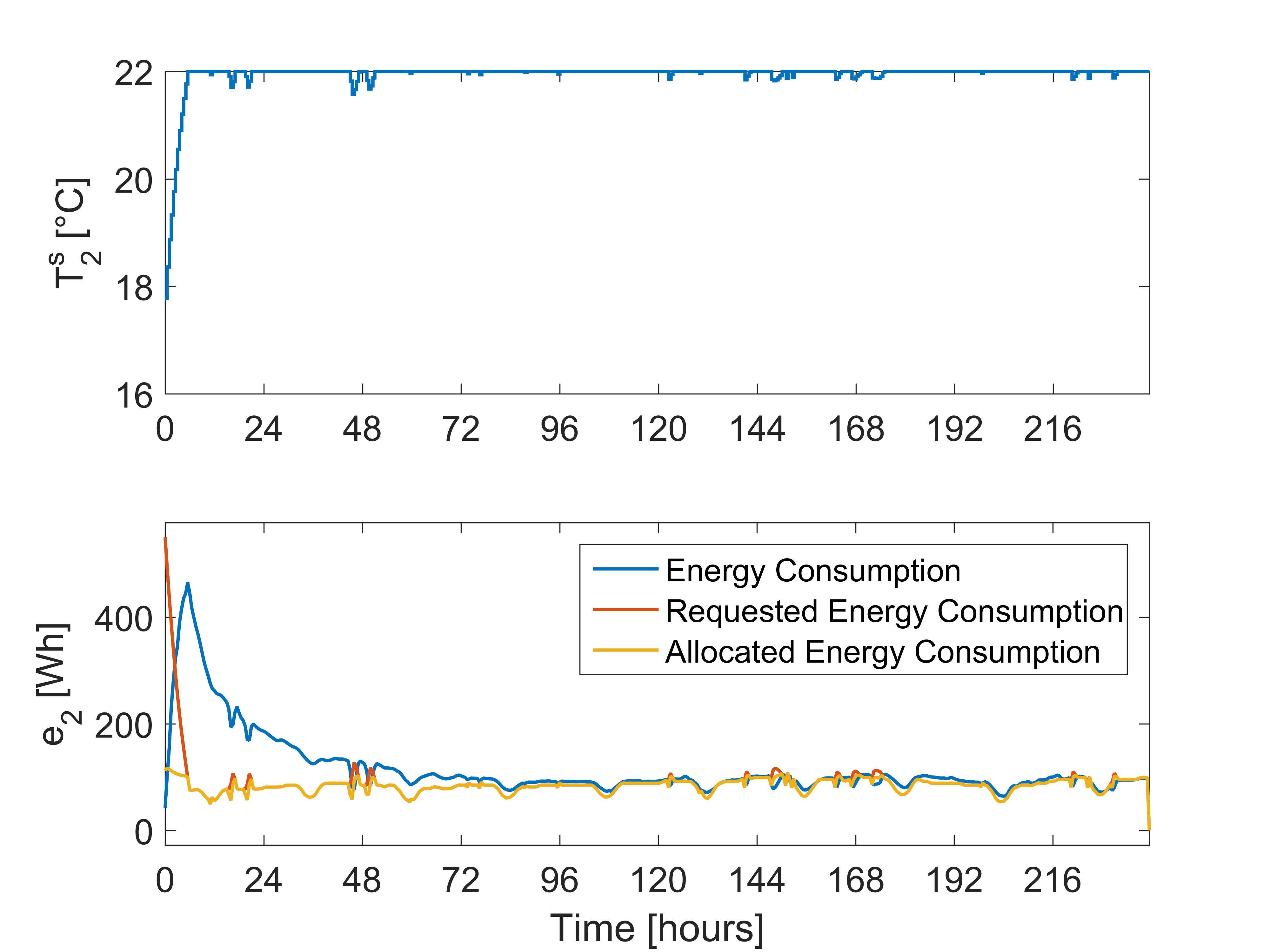}
        \caption{Karma-based Allocation, Unit 2}
        \label{fig:TsE_Z2_Karma_750}
    \end{subfigure}
    \hfil
    \begin{subfigure}[b]{0.49\linewidth}
        \centering
        \includegraphics[width=0.8\linewidth]{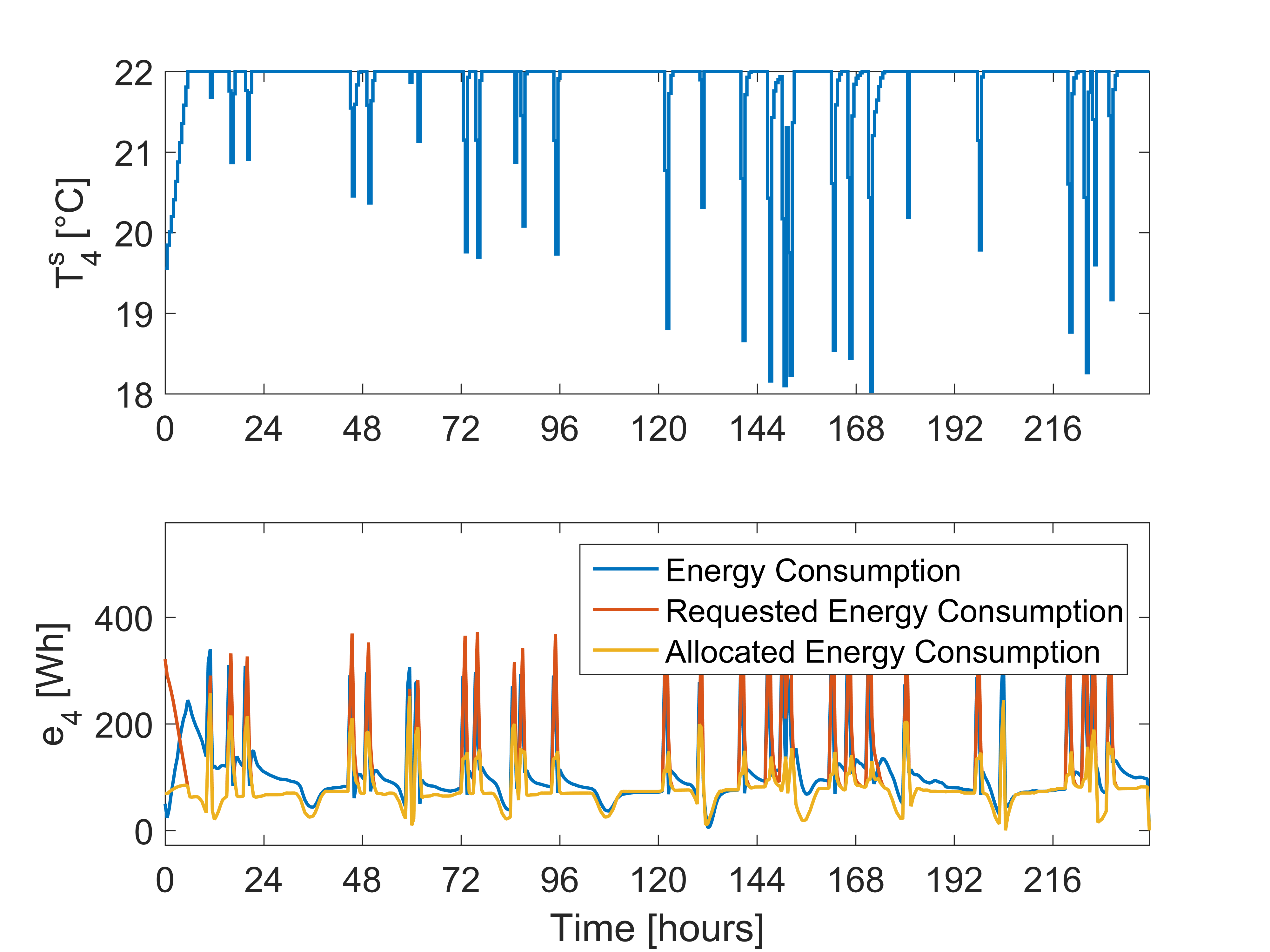}
        \caption{Karma-based Allocation, Unit 4}
        \label{fig:TsE_Z4_Karma_750}
    \end{subfigure}
    
    \caption{Results of Configuration 1 [Low Scarcity] with (1.1) proportional allocation baseline (top) versus (1.2) karma-based allocation (bottom), shown for Unit 2 (left) vs. Unit 4 (right).
    Top of each sub-figure: adjusted setpoints $\tilde T^s_i(t)$.
    Bottom of each sub-figure: measured energy consumption from IDA ICE versus requested and allocated energy ($e^r_i(t)$ and $e^a_i(t)$, respectively).}
    \label{fig:low-scarcity}
\end{figure*}

\begin{figure}[t]
    \centering
    \includegraphics[width=0.8\linewidth]{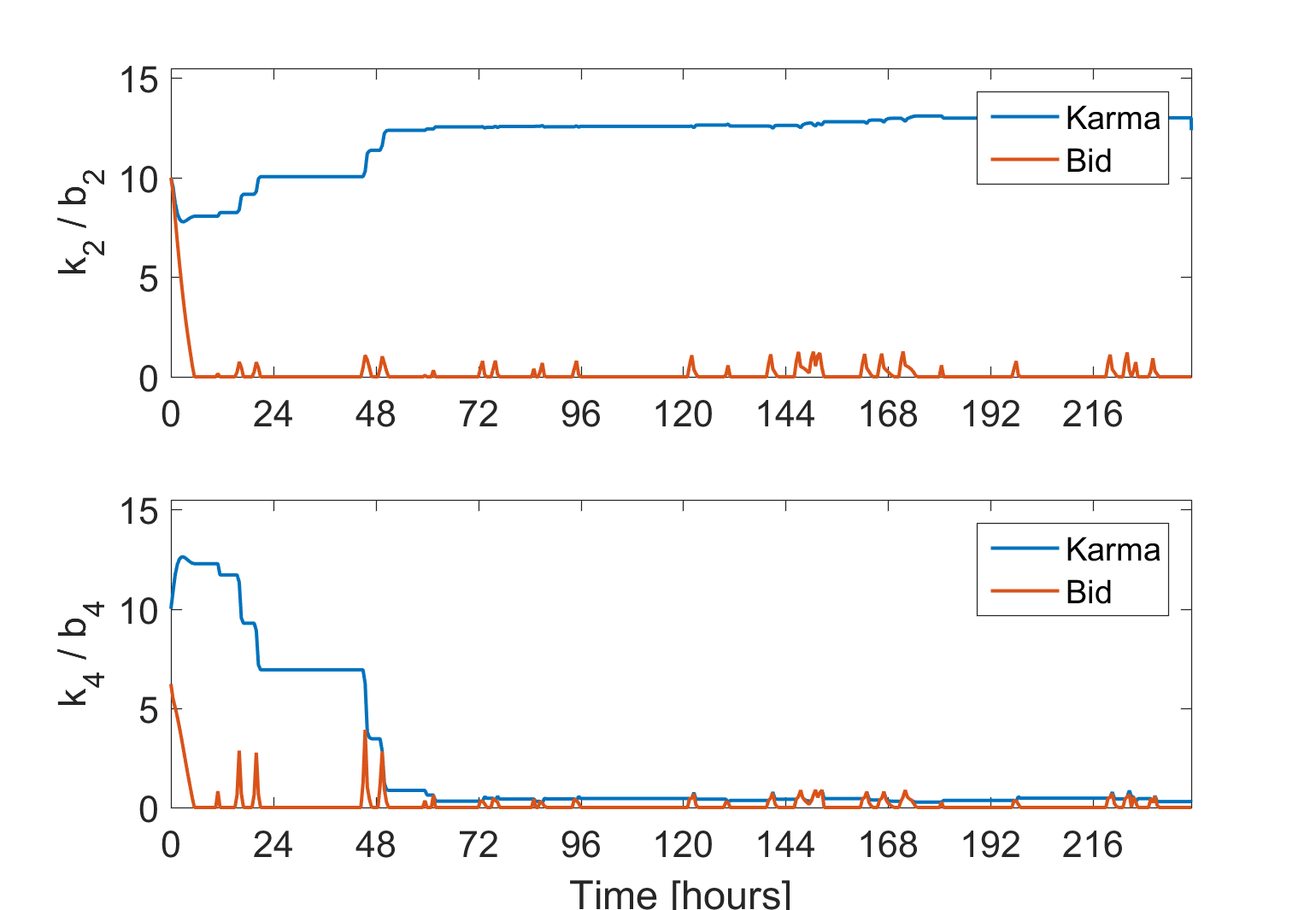}
    \caption{Karma $k_i(t)$ and bids $b_i(t)$ of Unit 2 (top) and Unit 4 (bottom) in the low-scarcity configuration.}
    \label{fig:kb_750}
\end{figure}

To provide further insight, Fig.~\ref{fig:low-scarcity} compares both the adjusted setpoints $\tilde T^s_i(t)$, as well as the space heating energy consumption (as truly measured vs. requested vs. allocated), between Unit 2 and Unit 4 in the low scarcity configuration, both under the baseline proportional and the karma-based allocations.
The proportional allocation baseline reduces the adjusted setpoint of Unit 2 as a consequence of Unit 4's window opening actions, leading to less space heating energy allocated and truly measured to Unit 2, cf. Fig.~\ref{fig:TsE_Z2_Karma_750}.
In contrast, Unit 2's adjusted setpoint is largely unaffected under the karma-based allocation, and it gets allocated its requested energy the majority of times, cf. Fig.~\ref{fig:TsE_Z2_Karma_750}.
Instead, Unit 4 is the one that experiences more drastic setpoint reductions and associated energy limits, cf. Fig.~\ref{fig:TsE_Z4_Karma_750}.
To explain this phenomenon, Fig.~\ref{fig:kb_750} compares the karma $k_i(t)$ and bids $b_i(t)$ of both units, and reveals that Unit 4's recurrent window opening and associated high energy requests gradually lead to depletion of that unit's karma, which henceforth limits its capacity to continue making high energy requests.


%% file: sections/conclusion.tex
\section{Conclusion}\label{sec:conclusion} 

This paper develops a co-simulation framework for validating a karma-based solution for budgeted space heating energy allocation, as a first step towards designing and conducting human-in-the-loop experiments in the KTH Live-In Lab.
The simulation results suggest that the karma-based scheme is effective in promoting sustainable heating behaviors, e.g., avoiding excessive window opening in the winter.
They also shed light on the required parameter tuning of the karma scheme.
Future work should investigate the sensitivity to these parameters and explore adaptive methods for selecting them.
Rather than following a fixed heuristic policy, an adaptive policy that learns and adapts to occupant behaviors could be preferred; however such adaptation risks de-stabilizing the scheme.
Finally, an important question is whether to (partially) expose occupants to the karma bidding or let the karma-scheme run autonomously in the background.

%% file: sections/bidding-policy.tex
\section{Derivation of Bidding Policy}
\label{sec:bidding-policy}

In~\cite{elokda2025vision}, it is shown that when karma budgets are relaxed to hold in a long-run average sense rather than in each time-step, bids enter the Langrangian of the associated individual optimization problems as an additive linear term, leading to optimal bids that are linear in payoffs, cf. \cite[Proposition~4]{elokda2025vision}.
While the specifics of the present setting differ from the one considered in~\cite{elokda2025vision}, both settings feature the same (relaxed) karma budget constraint, which motivates the linear structure adopted in our heuristic policy.
Namely, let us define $c_i(t) = e^r_i(t) - e^a_i(t)$ as the cost incurred by unit $i$ for receiving energy $e^a_i(t)$ when it requests $e^r_i(t)$.
Assuming that a) bids are linear in this cost, as motivated above, b) there is supply shortage, i.e., $\sum_j e^r_j(t) > e^b(t)$, and c) karma budget constraints do not bind gives rise to the bidding structure in Eq.~\eqref{eq:bidding-policy}:
\begin{align*}
&& b_i(t) &= \beta \left(e^r_i - e^a_i\right) =\beta \left(e^r_i - \frac{e^b}{\sum_j b_j} \, b_i\right) \\
&& &=\beta \left(e^r_i - \frac{e^b}{\beta \left(\sum_j e^r_j - e^b\right)} \, b_i\right) \\
\Leftrightarrow && b_i(t) &= \beta \left(1 - \frac{e^b}{\sum_j e^r_j}\right) e^r_i.
\end{align*}

%% file: main.bbl
\begin{thebibliography}{19}
\providecommand{\natexlab}[1]{#1}
\providecommand{\url}[1]{\texttt{#1}}
\providecommand{\urlprefix}{URL }
\expandafter\ifx\csname urlstyle\endcsname\relax
  \providecommand{\doi}[1]{doi:\discretionary{}{}{}#1}\else
  \providecommand{\doi}{doi:\discretionary{}{}{}\begingroup
  \urlstyle{rm}\Url}\fi

\bibitem[{Annaswamy et~al.(2023)Annaswamy, Johansson, and
  Pappas}]{annaswamy2024control}
Annaswamy, A.M., Johansson, K.H., et~al. (eds.) (2023).
\newblock \emph{Control for Societal-scale Challenges: Road Map 2030}.
\newblock IEEE Control Systems Society Publication.

\bibitem[{Dedecca et~al.(2025)Dedecca, Ansarin, Bene, Van~Delzen, Van~Nuffel,
  and Jagtenberg}]{dedecca2025increasing}
Dedecca, J.G., Ansarin, M., et~al. (2025).
\newblock Increasing flexibility in the eu energy system: Technologies and
  policies to enable the integration of renewable electricity sources.

\bibitem[{Drgo{\v{n}}a et~al.(2020)Drgo{\v{n}}a, Arroyo, Figueroa, Blum,
  Arendt, Kim, Oll{\'e}, Oravec, Wetter, Vrabie et~al.}]{drgovna2020all}
Drgo{\v{n}}a, J., Arroyo, J., et~al. (2020).
\newblock All you need to know about model predictive control for buildings.
\newblock \emph{Annual Reviews in Control}, 50, 190--232.

\bibitem[{Elokda et~al.(2024{\natexlab{a}})Elokda, Bolognani, Censi,
  D{\"o}rfler, and Frazzoli}]{elokda2024self}
Elokda, E., Bolognani, S., et~al. (2024{\natexlab{a}}).
\newblock A self-contained karma economy for the dynamic allocation of common
  resources.
\newblock \emph{Dynamic Games and Applications}, 14(3), 578--610.

\bibitem[{Elokda et~al.(2024{\natexlab{b}})Elokda, Bolognani, D{\"o}rfler, and
  Nax}]{elokda2024dynamic}
Elokda, E., Bolognani, S., et~al. (2024{\natexlab{b}}).
\newblock Dynamic resource allocation with karma: An experimental study.
\newblock \emph{arXiv preprint arXiv:2404.02687}.

\bibitem[{Elokda et~al.(2025{\natexlab{a}})Elokda, Cenedese, Zhang, Censi,
  Lygeros, Frazzoli, and D{\"o}rfler}]{elokda2025carma}
Elokda, E., Cenedese, C., et~al. (2025{\natexlab{a}}).
\newblock {CARMA}: Fair and efficient bottleneck congestion management via
  nontradable karma credits.
\newblock \emph{Transportation Science}, 59(2), 340--359.

\bibitem[{Elokda et~al.(2025{\natexlab{b}})Elokda, Censi, Frazzoli,
  D{\"o}rfler, and Bolognani}]{elokda2025vision}
Elokda, E., Censi, A., et~al. (2025{\natexlab{b}}).
\newblock A vision for trustworthy, fair, and efficient socio-technical control
  using karma economies.
\newblock \emph{Annual Reviews in Control}, 60, 101026.

\bibitem[{Fabi et~al.(2012)Fabi, Andersen, Corgnati, and
  Olesen}]{fabi2012occupants}
Fabi, V., Andersen, R.V., et~al. (2012).
\newblock Occupants' window opening behaviour: A literature review of factors
  influencing occupant behaviour and models.
\newblock \emph{Building and Environment}, 58, 188--198.

\bibitem[{Farjadnia et~al.(2026)Farjadnia, Fontan, Johansson, and
  Molinari}]{farjadnia2026assessing}
Farjadnia, M., Fontan, A., et~al. (2026).
\newblock Assessing the impact of occupant behavior on residential building
  performance: A case study of window operation.
\newblock \emph{Building and Environment}, 114552.

\bibitem[{Gr{\v{z}}ani{\'c} et~al.(2022)Gr{\v{z}}ani{\'c}, Capuder, Zhang, and
  Huang}]{grvzanic2022prosumers}
Gr{\v{z}}ani{\'c}, M., Capuder, T., et~al. (2022).
\newblock Prosumers as active market participants: A systematic review of
  evolution of opportunities, models and challenges.
\newblock \emph{Renewable and Sustainable Energy Reviews}, 154, 111859.

\bibitem[{{International Energy Agency}(2022)}]{IEA_Building}
{International Energy Agency} (2022).
\newblock Buildings: Energy system.
\newblock \url{https://www.iea.org/energy-system/buildings}.

\bibitem[{Kalamees(2004)}]{kalamees2004ida}
Kalamees, T. (2004).
\newblock {IDA ICE}: the simulation tool for making the whole building energy
  and ham analysis.
\newblock \emph{Annex}, 41, 12--14.

\bibitem[{Kelly et~al.(1998)Kelly, Maulloo, and Tan}]{kelly1998rate}
Kelly, F.P., Maulloo, A.K., et~al. (1998).
\newblock Rate control for communication networks: shadow prices, proportional
  fairness and stability.
\newblock \emph{Journal of the Operational Research society}, 49(3), 237--252.

\bibitem[{Kondziella and Bruckner(2016)}]{kondziella2016flexibility}
Kondziella, H. and Bruckner, T. (2016).
\newblock Flexibility requirements of renewable energy based electricity
  systems--a review of research results and methodologies.
\newblock \emph{Renewable and Sustainable Energy Reviews}, 53, 10--22.

\bibitem[{Moulin(2004)}]{moulin2004fair}
Moulin, H. (2004).
\newblock \emph{Fair division and collective welfare}.
\newblock MIT press.

\bibitem[{Nisan et~al.(2007)Nisan, Roughgarden, Tardos, and
  Vazirani}]{nisan2007algorithmic}
Nisan, N., Roughgarden, T., et~al. (2007).
\newblock \emph{Algorithmic Game Theory}.
\newblock Cambridge University Press, Cambridge.

\bibitem[{Palm and Thomas(2026)}]{palm2026americans}
Palm, M. and Thomas, A. (2026).
\newblock Americans meet congestion pricing: Reframing the equity debate around
  cbd tolling for auto-dependent countries.
\newblock \emph{Urban Studies}.

\bibitem[{Pedroso et~al.(2024)Pedroso, Agazzi, Heemels, and
  Salazar}]{pedroso2024fair}
Pedroso, L., Agazzi, A., et~al. (2024).
\newblock Fair artificial currency incentives in repeated weighted congestion
  games: Equity vs. equality.
\newblock In \emph{2024 IEEE 63rd Conference on Decision and Control (CDC)},
  954--959.

\bibitem[{Xu et~al.(2023)Xu, Yu, Sun, and Tam}]{xu2023critical}
Xu, X., Yu, H., et~al. (2023).
\newblock A critical review of occupant energy consumption behavior in
  buildings: How we got here, where we are, and where we are headed.
\newblock \emph{Renewable and Sustainable Energy Reviews}, 182, 113396.

\end{thebibliography}
